\documentclass[aps,prd,reprint,groupedaddress,amsmath,amssymb,superscriptaddress,nofootinbib]{revtex4-1}

\usepackage{graphicx}
\usepackage{dcolumn}
\usepackage{bm}
\usepackage[utf8]{inputenc}
\usepackage[T1]{fontenc}

\newcommand{\be}{\begin{equation}}
\newcommand{\ee}{\end{equation}}
\newcommand{\bear}{\begin{eqnarray}}
\newcommand{\eear}{\end{eqnarray}}

\newcommand{\p}{\prime}

\newcommand{\n}{\hat{n}}
\newcommand{\dv}{\delta v}

\newcommand{\cE}{{\cal E}}

\newcommand{\cD}{{\cal D}}

\newcommand{\cR}{{\cal R}}
\newcommand{\cT}{{\cal T}}
\newcommand{\cC}{{\cal C}}
\newcommand{\cA}{{\cal A}}

\newcommand{\cO}{{\cal O}}
\newcommand{\cQ}{{\cal Q}}

\newcommand{\rH}{{\rm H}}
\newcommand{\rN}{{\rm N}}

\newcommand{\rK}{{\rm K}}

\begin{document}

\title{On the  superradiance-tidal friction correspondence}

\author{Kostas Glampedakis}
\email[]{kostas@um.es}
\affiliation{Departamento de Fisica, Universidad de Murcia, Murcia, E-30100, Spain}
\affiliation{Theoretical Astrophysics, University of T\"{u}bingen, T\"{u}bingen, D-72076, Germany}

\author{Shasvath J. Kapadia}
\email[]{skapadia@uark.edu}
\affiliation{Department of Physics, University of Arkansas, Fayetteville, Arkansas 72701}

\author{Daniel Kennefick}
\email[]{danielk@uark.edu}
\affiliation{Department of Physics, University of Arkansas, Fayetteville, Arkansas 72701}
\affiliation{Arkansas Center for Space and Planetary Sciences, University of Arkansas, Fayetteville, Arkansas 72701}

\date{\today}

\begin{abstract}
Since the work of Hartle in the 1970s, and the subsequent development of the the Membrane Paradigm approach to 
black hole physics it has been widely accepted that superradiant scattering of gravitational waves bears strong similarities 
with the phenomenon of  ``tidal friction'' (well-known from Newtonian gravity) operating in binary systems of viscous material
bodies. In this paper we revisit the superradiance-tidal friction analogy within the context of ultracompact relativistic bodies. 
We advocate that as long as these bodies have non-zero viscosity they should undergo tidal friction that can be construed as a 
kind of superradiant scattering from the point of view of the dynamics of an orbiting test-body. In addition we consider the 
presence of anisotropic matter, which is required for at least some ultracompact bodies, if they are to sustain a radius very 
close to the gravitational radius. We find that the tidal friction/superradiance output is enhanced with increasing anisotropy 
and that strongly anisotropic systems exhibit an unconventional response to tidal and centrifugal forces. Finally, we make 
contact with the artificial system comprising a black hole with its horizon replaced by a mirror (sometimes used as a proxy for 
ultracompact material bodies) and discuss superradiance and tidal friction in relation to it.  

\end{abstract} 
  
\maketitle


\section{Introduction}
\label{sec:intro}

The Membrane paradigm \cite{membrane, damour} is a view of black holes which treats them not as regions of intense 
curvature in spacetime with a singularity at their center, but instead as two dimensional one-way membranes
situated at just about the position of the black hole's event horizon, but outside it so as to be visible to 
external observers. The membrane can be assigned a range of conventional physical properties not normally associated 
with black holes, including viscosity. This viewpoint enables calculations without the need for full relativity, and is 
frequently used for calculations which deal with the electromagnetic properties of the black hole. One striking feature 
of the membrane paradigm is that it also provides a way of understanding an important result in black hole physics due to Hartle 
in the 1970s \cite{hartle73,hartle74} in which a black hole in a binary system can transfer its rotational angular momentum to the
orbital angular momentum of a satellite in a manner analogous to tidal friction in material bodies. In the membrane paradigm this 
can be understood as the result of the tides raised in the event horizon membrane by the orbiting body. If the orbital period of 
the satellite is not identical to the rotational period of the black hole, then the effective viscosity of the event horizon will 
prevent the tidal bulges from remaining under the satellite and the resulting gravitational torque will result in the transfer of 
angular momentum. In the most typical case, where the orbital frequency of the satellite is smaller than the rotational frequency
of the black hole, this involves braking the black hole and accelerating the satellite, permitting it to draw further away from the 
central body.

Hartle's approach is, however, rarely used by physicists in calculating the angular momentum transfer between the bodies in the case 
of extreme-mass ratio black hole binaries. Rather the Teukolsky formalism (or a similar formalism) is used instead to calculate the 
gravitational waves emitted towards infinity by the system. The event horizon can be treated as an internal infinity in the system and 
the flux of gravitational waves emitted by the orbiting body (treated as a perturbation of the central black hole's spacetime) which 
reach that infinity can be calculated. It is generally found (for prograde orbits where the body orbits in the same sense as the black
hole rotates, but again with a greater angular frequency) that the flux of energy and angular momentum reaching the horizon is negative. 
Thus the black hole loses mass and angular momentum by this interaction, while the small body gains the same amount.
This phenomenon is known as superradiance, the implication being that waves are superradiantly scattered back by the black hole, 
the emitting body thus receiving more energy than it emitted. Although superradiance is today associated with rotating black holes, 
and therefore with both the presence of an event horizon and an ergoregion, in fact it requires neither. The phenomenon was first 
discussed by Zel'dovich in the context of electromagnetic waves scattering off a non-relativistic cylinder
\cite{Zel'Dovich71,Zel'Dovich72,bs98,RichartzSaa13}.

While it is sometimes observed that superradiant scattering and Hartle-like tidal friction are the same phenomenon, this has not been shown explicitly in full relativity. Moreover it is also of interest to ask whether they are physically the same effect, rather than a mere analogy. Since it is difficult to compare black holes to normal material bodies, it  is convenient to consider a hypothetical material 
body with a field like that of a black hole, a body even more compact than a neutron star. Since we know little about the structure
of such objects, if they exist, a common stand-in is the so-called mirror Kerr spacetime, which is simply a Kerr black hole with a perfectly reflecting mirror placed just in front of the event horizon. From the point of view of radial motion there is little difference between a 
perfect mirror and a perfectly transparent central body, so the mirror Kerr system is a convenient surrogate for an ultracompact star.

Investigating this mirror-Kerr spacetime Richartz et al.~\cite{richartzetal} have shown recently that superradiance disappears 
with the introduction of the mirror. This is not very surprising, since by definition a perfect mirror will produce a reflected wave
equal in amplitude to the incident wave, rather than one with greater amplitude. Richartz et al. demonstrated their result for a scalar
field, so in this paper we perform their calculation for a tensor field to demonstrate that the result does also work for gravitational waves. 
A natural interpretation of the result would be that hypercompact stars whose material was transparent (or nearly so) to gravitational waves should not exhibit superradiance in a binary system, and therefore, if superradiance and tidal friction are equivalent, should not undergo tidal friction either. But we would normally expect to observe tidal friction in any binary system consisting of material bodies. We are interested in finding whether both superradiance and tidal friction can be found in the case of purely material objects (without an event horizon) and whether they are still clearly two ways of describing the same physical effect.

Tidal friction should occur in the ultracompact material system if the central body, whatever its nature, has viscosity. So we address
the role of viscosity in the binary system, whether viewed within the superradiance or the tidal friction framework. It has been known 
since the 1970s that viscosity plays a central role in the absorption of gravitational wave energy by a fluid body \cite{Esposito71,PapadopoulosEsposito85}. 
The introduction of viscosity will permit us to maintain the identification of tidal friction 
and superradiance in ultra-relativistic objects. If viscosity goes to zero (as in the mirror-Kerr example, which presumes the central 
body is transparent to gravitational waves) then both tidal friction and superradiance goes to zero. If viscosity remains, then both
effects remain operative.

In Section~\ref{sec:hartle} we will review the standard Newtonian and post-Newtonian theories of both Hartle-like tidal 
friction and superradiance and show that, to lower post-Newtonian order, they produce identical results.

Since we are interested in hypercompact objects we take note of results which suggest that such objects cannot support themselves without some anisotropy in their internal pressure. There has been, to our knowledge, no study of how pressure anisotropy would 
affect tidal interactions, even in the Newtonian case, so we begin by calculating tidal friction where such anisotropy exists. 

In Section~\ref{sec:torque} we show that the anisotropy can amplify the conventional tidal friction result, so that hypercompact objects, 
if they exist and are anisotropic, may have augmented tidal coupling with an orbiting satellite. An interesting result is that bodies with significant anisotropic pressure, when they are exposed to a tidal force, form a tidal bulge which is nearly orthogonal (in the direction
of its major axis) with respect to the bulge formed in isotropic systems. Moreover, these systems are characterized by a negative
tidal Love number and a prolate (rather than oblate) shape caused by rotation.

In Section~\ref{sec:newton} we study the relation between superradiance and tidal friction in more detail, and identify 
an effective viscosity and an effective tidal Love number for black holes.

In Section~\ref{sec:objects} we carefully study the extent to which superradiance can be said to occur in bodies with no event horizon.
Moreover, we speculate about the viscosity of ultracompact objects and about the possibility of having floating orbits around them.

In Section~\ref{sec:mirror} we study the mirror-Kerr spacetime. There are two key issues discussed. One is the introduction of an 
imperfect mirror, designed to model a viscous central body which can absorb gravitational wave energy, though not perfectly as in 
the black hole case. The other is to take account of the potential barrier surrounding the black hole-like body, which can set up a cavity
within which waves reflect back and forth between the mirror and the potential barrier  as discussed in \cite{vilenkin78}. 
We show that superradiance and the so-called ergoregion instability are both possible in this system, depending on the reflectivity of 
the mirror (which is probably related to the viscosity of the material system). We conclude in Section~\ref{sec:conclusions} with a 
discussion of the interpretation of our various results.

\section{Superradiance and tidal torque on black holes}
\label{sec:hartle}

In the 1970s Hartle~\cite{hartle73, hartle74} showed that black holes could transfer their spin angular momentum to an orbiting body 
by a process remarkably similar to tidal friction in material systems such as the Earth-Moon. In this Section we provide an outline of
Hartle's tidal theory and its connections to the phenomenon of black hole superradiance.  As such, our discussion here adds little new information   
on the subject but, nevertheless, it serves to set the scene for the main analysis of this paper.

Hartle derived a formula for the torque exerted 
on a Kerr black hole, of mass $M$ and spin angular momentum $J = a M^2$, by a test body of mass $\mu \ll M$ in a circular orbit around the black hole
(for simplicity we assume an equatorial orbit with angular frequency $\Omega_{\rm orb}$). The orbit is assumed to lie in the weak-field 
region with a radius $b \gg M$ and as a consequence the orbital velocity is itself small, $v^2 = M/b \ll 1$.  

In addition, Hartle's analysis assumed a slowly spinning black hole in the sense that only leading order terms with respect 
to the Kerr parameter $a$ are retained. Then the angular frequency of the hole's horizon can be approximated as,
\be
\Omega_\rH = \frac{a}{2r_{+}} \approx \frac{a}{4M}
\ee
where $r_{+} = M + M\sqrt{1-a^2} \approx 2M$ is the event horizon radius in standard Boyer-Lindquist coordinates. 
The two characteristic frequencies of the system, $\Omega_{\rm orb}$ and $\Omega_\rH$, are assumed to obey
\be
M \Omega_{\rm orb} \ll M\Omega_\rH \ll 1 
\label{cond1}
\ee
This ordering implies that within Hartle's model we are not allowed to take the Schwarzschild limit 
$a \to 0$. We are allowed to make $\Omega_{\rm orb}$ arbitrarily small so that we have a quasi-stationary test-body 
with respect to the hole (as, in fact, Hartle did). 

Hartle's formula for the tidal torque exerted on the black hole is given by~\cite{hartle73, hartle74}:
\be
\dot{J} = -\frac{8}{5} \frac{\mu^2 M^3}{b^6} J 
\label{hartle1}
\ee
where an overdot represents a time derivative. As we have mentioned, this expression is accurate to leading order 
with respect to both $J$ and $v$. An equivalent form of (\ref{hartle1}) is,
\be
\dot{J} = -\frac{32}{5} \frac{\mu^2 M^6}{b^6} \Omega_\rH
\label{hartle2}
\ee
This result implies that a slowly-rotating Kerr hole is always spun down under the action of the torque caused by a distant 
orbiting body (provided~(\ref{cond1}) holds). Global conservation of angular momentum then requires that the rotational energy removed 
from the hole is gained by the orbiting body. 

This gain of energy by the small body at the expense of the black hole bears an obvious resemblance to 
the situation where energy is removed by a black hole via superradiant scattering of gravitational
waves~\cite{chandrabook}. It can be shown that this is more than a mere similarity and we will investigate the extent to
which Hartle's spin-down effect is physically indistinguishable from superradiance. 
 
This equivalence can be demonstrated by considering the gravitational wave energy flux at the horizon as calculated
by a post-Newtonian approximation of the Teukolsky black hole formalism~\cite{teuk73}. This arduous calculation is described in 
Ref.~\cite{chapter}; the final result for the energy flux at the horizon produced by a test body in a circular equatorial orbit is:
\bear
&& \dot{E}_\rH = \dot{E}_\rN v^5 \Bigg [ -\frac{1}{4} a \left (  1 + 3a^2 \right ) - a\left ( 1 + \frac{33}{16} a^2
\right ) v^2 
\nonumber \\
\nonumber \\
&& \qquad + \left ( 1 + {\cal O}(a^2) \right ) v^3 + {\cal O}(v^4) \Bigg ]
\label{PNkerr}
\eear
where 
\be
\dot{E}_\rN = \frac{32}{5} \frac{\mu^2}{M^2} v^{10}
\ee
The analysis leading to this result only assumes $v \ll 1$, without placing any restriction on the Kerr parameter, $0 \leq a <  1$. 
The negative sign of the leading (and higher) order spin-dependent terms in (\ref{PNkerr}) is associated with superradiance 
since it represents an energy loss for the black hole. 

The situation is quite different for a Schwarzschild black hole. According to (\ref{PNkerr}), for $a=0$ the horizon flux first appears at a higher post-Newtonian order and it is positive, that is, it represents energy removed from the orbiting body and absorbed by the black hole.

We shall now derive a spindown formula based on the horizon flux (\ref{PNkerr}). 
For a circular orbit the angular momentum flux down the black hole horizon is given by 
(see for instance~\cite{kennefick98})
\be
\dot{J}_\rH = \frac{1}{\Omega_{\rm orb}} \dot{E}_\rH 
\label{JandE}
\ee
Expanding the angular frequency in a post-Newtonian series (see Ref.~\cite{chapter}) and combining 
it with the above flux formula we obtain:
\bear
&& \dot{J}_\rH = -\frac{32}{5}  \frac{\mu^2  M^6}{b^6} \Omega_\rH
\Bigg [  1 + 3a^2  - \left ( 4 + \frac{33}{4} a^2
\right ) v^2 
\nonumber \\
&& \qquad  - \left ( \frac{4}{a} + {\cal O}(a^2) \right ) v^3 + {\cal O}(v^4) \Bigg ]
\label{Jdot1}
\eear
A similar result of even higher precision has been derived in Refs.~\cite{taylor08,CPY13}.  
If we wish to truncate this series at ${\cal O}(v^2)$ order (or lower) then we need to make
sure that in the slow-rotation limit $a \ll 1$ we also have
$a \gg v = (M/b)^{1/2}$;
it is easy to see that this condition implies (\ref{cond1}).

The overall negative sign in the $\dot{J}_\rH$ flux (=spindown) is a direct consequence of 
superradiance in the gravitational wave energy flux $\dot{E}_\rH$. 
At the same time, and at leading order with respect to $a$ and $v$, $\dot{J}_\rH$ is identical to the Hartle torque
(\ref{hartle2}).
Therefore the spindown described by (\ref{hartle2}) can be fully attributed to superradiance. In other words
Hartle's ``tidal friction'' is simply superradiance. 

It is also clear that  (\ref{Jdot1}) can be viewed as an extension of Hartle's formula for an arbitrary
spin $a$ (note that similar formulae have been derived in the past, see Refs.~\cite{chrzanowski76, poisson04}).
In particular, the angular momentum flux at the horizon of a Schwarzschild black hole (more precisely of a
hole with $\Omega_\rH \ll \Omega_{\rm orb}$) is
\be
\dot{J}_\rH \approx \frac{32}{5} \frac{\mu^2 M^6}{b^6} \Omega_{\rm orb} 
\label{Jdot2}
\ee
and represents a spin-up torque. It can be seen that this expression displays the same dependence with respect to $M$ 
and $b$,  as well as the same numerical prefactor $32/5$ (albeit with opposite sign), as that of the slow rotation result (\ref{hartle2})
(or (\ref{Jdot1})).

The two torque results we have discussed so far, that is,  the torque exerted on the black hole in systems with $M\Omega_{\rm orb} \ll 1$ 
and $M\Omega_\rH \ll 1$ and with $\Omega_\rH/\Omega_{\rm orb} \gg 1$ (eqn.~(\ref{hartle2})) or $\Omega_\rH / \Omega_{\rm orb} \ll 1 $
(eqn.~(\ref{Jdot2})), can be combined into a `global' torque formula:
\be
\dot{J}_\rH \approx \frac{32}{5} \frac{\mu^2 M^6}{b^6} ( \Omega_{\rm orb} - \Omega_\rH )
\label{Jdot4}
\ee
This is valid for any ratio $\Omega_\rH / \Omega_{\rm orb}$ and at leading order with respect to the small parameter 
$ | M(\Omega_{\rm orb} -\Omega_\rH) | \ll 1$. A key property of (\ref{Jdot4}) is that it predicts a zero torque on the black hole
in the special case of a black hole-synchronous orbit, $\Omega_{\rm orb} = \Omega_\rH$. In fact, this is generally true for all circular
equatorial orbits in Kerr spacetime -- this can be verified by a simple inspection of the fully relativistic formula for $\dot{J}_\rH$
as calculated with the help of the Teukolsky formalism (see Ref.~\cite{kennefick98} for details).


\subsection{The nature of tidal bulges on the event horizon}

In a binary system, if the body upon which tidal bulges are raised rotates with respect to the source of the tidal gravitational 
field then the viscosity of the body causes the bulges to attempt to rotate with it, even though they ``ought'' to remain directly 
under the body which is the source of the tidal field.
The upshot is a tidal lag, in which the bulges take up a position some few degrees away from their ``natural'' position. This introduces 
a torque acting on the body because the pull of the source is no longer 
purely radial, but also tangential. This
process transfers angular momentum from the rotating body to the orbiting
body (in the most familar cases), with some additional dissipation of energy, for instance by heating.
In cases where the orbiting body actually has a larger angular frequency
than that of the rotating body (as in the case of Phobos, the moon of Mars),
then a tidal lead will occur and angular momentum will be transferred the
other way, from the satellite, operating to speed up rather than retard the
rotation of the planet. This process is speculated to explain the moonlessness
of the planet Venus, whose extremely slow rotation would cause any satellite
to lose orbital angular momentum and eventually crash into the planet below.
In a similar process, we may expect nearly Schwarzschild black holes to more
quickly draw and swallow orbiting objects than would rapidly rotating black
holes.

At first glance it is not at all apparent that there should be an analogous process operating in black holes. The mass of the black
hole is concentrated at a point (at least classically) and it seems counter-intuitive that the event horizon should have a viscosity.
Nevertheless Hartle showed that an interaction analogous to tidal friction does take place between the event horizon of a black
hole and an orbiting body. This insight led to the development of a different view of the black hole, known as the membrane paradigm~\cite{membrane}, 
in which the event horizon is conceived of as a surface which interacts physically with the outside Universe with certain properties determined 
by the mass and spin of the black hole. In this viewpoint the surface has an effective viscosity which explains the dissipation connected
with the exchange of angular momentum between the black hole and the orbiting body. 

It is worth noting, however, that the analogy is not perfect where the actual shape of the tidal distortion of the event horizon is concerned. 
Several studies have shown that the tidal bulge raised in an event horizon is \emph{not} in the position one would expect from the case of 
material bodies \cite{hartle74, FangLovelace05, VPM11}. 

Fang \& Lovelace \cite{FangLovelace05} and Vega, Poisson \& Massey \cite{VPM11} argue that this is the 
result of the teleological nature of the horizon, based on Hartle's \cite{hartle74} original 
point that the boundary condition which defines 
the location of the horizon is imposed at future null infinity and
thus has an inherently different causality to that which applies to
the surface of a material body.
Although the actual position of the tidal bulge is different in the black hole case, it produces the same effects, in terms of energy and 
angular momentum transfer, as does a bulge in a material body.


\section{Tidal torque on Newtonian anisotropic stars}
\label{sec:torque}

\subsection{Why anisotropic stars?}

The second stage of our analysis of the superradiance-tidal friction correspondence requires a discussion of the
tidal torque exerted on material fluid bodies. The classic Newtonian tidal theory was first formulated in the 
late 19th century by Darwin \cite{darwin} and is still in use today. One of the modern discussions of Darwin's theory is provided in a 
recent paper by Poisson~\cite{poisson09}, in the context of tidally deformed black holes and their similarities with Newtonian tides. 

Our own discussion of tidal friction in this Section relies heavily on Poisson's analysis, differing only
in that we consider a Newtonian star made of \emph{anisotropic} matter (this property is encoded in the presence
of two scalar pressures, radial and tangential, as described below). 

The motivation for choosing to study anisotropic stars is this: from the known solutions of compact relativistic stars
it is clear that these objects cannot be made arbitrarily compact (i.e. the stellar radius $R_\star$ approaching the Schwarzschild
radius $2M_\star$) unless their matter is highly anisotropic. For instance, this is the case for uniform density spheres where only the
anisotropic Lemaitre solution can approach the `black hole' limit $R_\star \to 2M_\star $ \cite{lemait,bl74}. Another type of ultracompact 
objects, the so-called gravastars, are also known to require a layer of anisotropic pressure, see \cite{cattoen05}. 

Since only relativistic hypercompact stars can closely mimic black holes, it makes sense, if these bodies are likely
to exhibit pressure anisotropies, to understand how such anisotropies affect tidal deformation, albeit in Newtonian theory. 
As we shall see, the presence of anisotropy in matter leads to some rather interesting behaviour. 


\subsection{The unperturbed star}
\label{sec:bg}

To begin our analysis we first need to consider the unperturbed spherically symmetric star. The key stellar
parameters are the mass $M_\star$, radius $R_\star$, and pressures $p_r$ (radial) and $p_t$ (tangential). The star
is assumed to have a uniform density
\be
\rho = \frac{3M_\star}{4\pi R_\star^3}
\ee
It is also convenient to define the anisotropy parameter
\be
\sigma = p_r -p_t
\ee 
The hydrostatic equilibrium of the system is described by (we use the index $0$ to identify 
background parameters)
\be
p_{r0}^\p = -\rho \Phi_0^\prime  - \frac{2\sigma_0}{r}
\label{euler0}
\ee 
where a prime stands for a radial derivative. The gravitational potential $\Phi_0$ can be calculated in a standard way:
\be
\Phi_0 (r) = -\frac{M_\star}{2 R_\star} \left ( 3 - \frac{r^2}{R_\star^2} \right )
\label{phi0}
\ee

The stellar model would be complete if equations of state (that is, functional relations between the pressures and the density) were available but no
such physically-motivated relations are known for anisotropic matter. The only viable alternative is to use a convenient parametric relation between
$p_{t0}, p_{r0}$ and $\rho$ that can smoothly vary from isotropy to extreme anisotropy. One such parametrization is given in Ref.~\cite{bl74}. 
By taking the Newtonian limit of the general relativistic solution derived in that paper we have the following relation:
\be
p_{t0} (r) = p_{r0}(r) + \frac{4\pi r^2}{3} \rho^2 \left ( \frac{1}{2} -Q \right ) 
\label{pt1}
\ee
The parameter $Q$ controls the degree of anisotropy, allowing a smooth transition from isotropic stars ($Q=1/2$, the relativistic
Schwarzschild solution~\cite{MTW}) to the most extreme case of anisotropy, the so-called `Lemaitre vault' solution, 
$p_{r0}=0$ and $Q=0$~\cite{lemait,bl74}. The \textit{minimum} allowed stellar radius for a given $Q$ is \cite{bl74}:
\be
R_\star^{\rm min} = 2 M_\star \left ( 1 -3^{-1/Q} \right )^{-1}
\ee
which shows that $Q =0$ ($Q=1/2$) represents the most (the least) compact object.


\subsection{The perturbed star}

We now consider the previous anisotropic star, with rotation and under the action of an imposed
\emph{quadrupolar} tidal field (the tidal field source will be later specified as that of an orbiting test-body). 
Working in the stellar rest frame and using Cartesian coordinates $x^i$ we can express the tidal field
in terms of the time-dependent symmetric and trace-free (STF) tensor $\cE_{ij} (t)$. The induced tidal potential is
\be
\Phi_{\rm tidal} = \frac{1}{2} \cE_{ij} x^i x^j
\ee
Under the action of this field a stellar scalar parameter $X$ can be decomposed into a background
spherically symmetric piece and a quadrupolar perturbation, i.e.
\be
X = X_0 (r)+ \delta X =  X_0 (r) + X_{ij}x^i x^j
\ee
with $X_{ij} (t)$ being itself a STF tensor. 

Stellar rotation, with angular frequency vector $\Omega \hat{z}^i$ (a `hat' labels a unit vector), 
leads to a centrifugal potential which can also be decomposed in a similar way:
\be
V_{\rm cf} = \frac{1}{2} \Omega^2 ( x^2 + y^2) =  \Omega^2 \left ( \frac{1}{3} r^2 + \frac{1}{2} C_{ij} x^i x^j \right )
\ee
where $C_{ij}= \delta_{ij} /3 - \hat{z}_i \hat{z}_j$. 
The star is assumed to be slowly rotating, in the sense that $\Omega \ll \Omega_{\rm K}$,
where $\Omega_{\rm K} \sim (GM_\star/R^3_\star)^{1/2}$ is the mass-shedding spin limit\footnote{Strictly speaking, the spherically 
symmetric portion of $V_{\rm cf}$ should have been included in the calculation of the structure of the unperturbed star. However, it turns out 
that the rotational correction to the equations of Section~\ref{sec:bg} can be ignored at the level of precision of the tidal deformation calculation.}. 

The tidal field deforms the shape of the star and produces a velocity field $\dv^i$ which is superimposed on
the bulk rigid-body rotation. This velocity can be decomposed as~\cite{poisson09}
\be
\dv_i = ( r^2 + \gamma R^2_\star ) V_{ij} x^j + \beta V_{kl} x^k x^l x_i
\label{dv1}
\ee
where $\gamma,~\beta$ are dimensionless constants and $V_{ij}(t)$ is another STF tensor. 

The perturbed stellar surface can be parametrised with the help of the deformation tensor $e_{ij}(t)$ as:
\be
R = R_\star ( 1+ e_{ij} \n^i \n^j )
\label{R1}
\ee 
where $\n^i = x^i/r$ is the unit radial vector. The departure from sphericity is accompanied by an
additional `body' contribution $\Phi_{\rm body}$ to the total gravitational potential. We thus have
\be
\delta \Phi = \Phi_{\rm tidal} + \Phi_{\rm body} = \frac{1}{2} \cE_{ij} x^i x^j + \Phi_{ij} x^i x^j
\ee
The body potential can be calculated in the standard way~\cite{poisson09},
\be
\Phi_{ij} = -\frac{3}{5} \frac{M_\star}{R^3_\star} e_{ij}
\ee 
The velocity perturbation at the surface can be written as
\be
\n_j \dv^j (R_\star) = \partial_t R
\label{surf1}
\ee
Moreover we can assume an incompressible flow, $\delta \rho =0 \to \partial_j \dv^j = 0$. From these equations we can 
obtain $\beta=-2/5$ and
 \be
\partial_t e_{ij} = ( 1 + \beta + \gamma) R^2_\star  V_{ij}
\label{eV1}
\ee

Up to this point the perturbation analysis has been identical to that of isotropic fluid stars hence it is not surprising
that the previous equations match those of Ref.~\cite{poisson09}. However, the remaining equations to be discussed, the Euler
equation and the surface boundary condition, do depend on the presence of anisotropic matter.  

The linearised Euler equation describing the hydrodynamics of an incompressible viscous fluid with anisotropic pressure can be written
as (see the Appendix for a derivation)
\begin{multline}
\partial_t \dv_i  + 2\Omega \epsilon_{ijk} \hat{z}^j \dv^k = 
  \partial_i \left ( \frac{1}{2} \Omega^2 C_{jk} x^j x^k -\delta \Phi \right ) 
\\
-\frac{1}{\rho} \partial_i \delta p_t - \n_i \left (  \frac{1}{\rho} \n^j\partial_j \delta \sigma 
+ \frac{2}{r} \frac{\delta \sigma}{\rho} \right ) + \nu \nabla^2 \dv_i 
\label{fulleuler}
\end{multline}
This equation features the (perturbed) Coriolis and centrifugal forces, pressure and gravitational forces, 
and a viscous force with (uniform) shear viscosity coefficient $\nu$.

As far as the tidal interaction problem is concerned the full Euler equation~(\ref{fulleuler}) can be greatly
simplified if the star is slowly rotating ($\Omega\ll \Omega_\rK$) and the tidal field is slowly-varying (more precisely, for the 
particular case of tides raised by an orbiting body of angular frequency $\Omega_{\rm orb}$ we assume $|\Omega-\Omega_{\rm orb}| \ll \Omega_\rK$).
Under this double assumption we can easily show that the first two inertial terms in (\ref{fulleuler}) are negligible with respect to
the tidal force (whereas the centrifugal force is not), see also~\cite{poisson09}. The viscosity force is also assumed to be small; 
the mathematical requirement is $\tau_\nu |\Omega-\Omega_{\rm orb}| \ll 1$ ($\tau_\nu$ is the viscous delay timescale, see eqn.~(\ref{tvisc}) below 
for its definition). In essence this is a statement of a \emph{small phase-lag} between the instantaneous position of the orbiting body and the
direction of the raised tidal bulge on the star. It should be also pointed out that the viscosity term can even 
be much smaller than the other terms in the Euler equation (indeed it can be exactly zero) but, nevertheless, it must be retained 
as the \emph{leading} dissipative term. After all, this is the term responsible for the system's `tidal friction'.

Adopting the above assumptions the Euler equation can be approximated by the simpler, amputated expression:
\begin{multline}
\partial_i \left (\, \frac{1}{2} \Omega^2 C_{jk} x^j x^k - \frac{\delta p_t}{\rho} -\delta \Phi \, \right )
\\
- \n_i \left \{\,   \n^j\partial_j \left ( \frac{ \delta \sigma}{\rho}  \right )
+ \frac{2}{r} \frac{\delta \sigma}{\rho} \, \right \}  + \nu \nabla^2 \dv_i = 0 
\label{euler1}
\end{multline}

As a final equation we need to formulate a boundary condition of a vanishing force at the stellar surface. 
This can be done by defining the traction vector from the fluid stress-energy tensor, $t^i = T^{ij} \n_j$. 
The desired surface boundary condition is $t^i=0$ which leads to
\be
p_r \n_i  - \rho \nu (\partial_j  \dv_i + \partial_i \dv_j) \n^j = 0
\label{surf}
\ee


\subsection{Calculating the tidal deformation}

The torque exerted on the star by the tidal field can be computed once the deformation tensor $e_{ij}$
is available. The calculation of this parameter is the subject of this subsection.

The total surface pressure can be approximated as (using the fact that $p_{r0} (R_\star) = 0$),
\be
p_r (R) \approx  R^2_\star  \n_i \n_j  \left (\, p_r^{ij}  + \frac{p_{r0}^\p (R_\star)}{R_\star} e^{ij} \, \right )
\ee
Using this and eqn.~(\ref{dv1}) in the surface condition (\ref{surf}),
\begin{multline}
2\rho \nu \left [ ( 2 +\beta+\gamma)  V_{jk} \n^k + (1+2\beta) V_{kl} \n^k \n^l \n_j \right ]
\\
 =  \n_j \n_k \n_l \left ( p^{kl}_r + \frac{p_{r0}^\p(R_\star)}{R_\star} e^{kl} \right )
\end{multline}
The various terms in this equation can be collected in two independent groups, leading to $\gamma = -(2+\beta)= -8/5$
and
\be
p_r^{ij} = -\frac{p_{r0}^\p (R_\star)}{R_\star} e^{ij} - \frac{2\rho \nu}{R_\star^2} (1+2\beta) \partial_t e^{ij}
\ee
In the final step of this calculation we need to use the Euler equation (\ref{euler1}); this should eventually
become a differential equation for $e_{ij}$. Using the previous results,
\begin{multline}
x_l \left (\, \frac{1}{2} \Omega^2 C^{jl} -\frac{p_t^{jl}}{\rho} -\Phi^{jl} -\frac{1}{2} \cE^{jl}\, \right )
\\
- 2 x_l \n_j \n^k \frac{\sigma_{kl}}{\rho} = - \nu (5 + 2\beta) V^{jl} x_l
\label{euler2}
\end{multline}
After collecting terms of similar structure we find
\be
\sigma_{ij}= 0 \quad \to \quad p_r^{ij} = p_t^{ij}
\ee
and
\begin{multline}
\frac{\nu}{R^2_\star} (3-2\beta) \partial_t e_{ij} - \left (\, \frac{p_{r0}^\p (R_\star)}{\rho R_\star}
+ \frac{3}{5} \frac{M_\star}{R^3_\star} \,  \right ) e_{ij}  
\\
= \frac{1}{2} \left ( \Omega^2 C_{ij} - \cE_{ij} \right )
\end{multline}
The pressure gradient can be eliminated with the help of the hydrostatic equilibrium equation
(\ref{euler0}). Then, after using (\ref{pt1}) and some trivial algebra, we arrive at:
\be
\frac{\tau_0}{\cA} \partial_t e_{ij} + e_{ij} 
= \frac{5}{4 \cA} \frac{R^3_\star}{M_\star}\left ( \Omega^2 C_{ij} - \cE_{ij} \right )
\label{e1}
\ee
where we have introduced the `viscous time delay' appearing in the tidal theory of isotropic stars~\cite{poisson09}:
\be
\tau_0 = \frac{19}{2} \frac{\nu R_\star}{M_\star}
\label{tau0}
\ee
The only difference between our result (\ref{e1}) and the corresponding formula for isotropic stars is encoded
in the anisotropy factor,
\be
\cA \equiv 5 Q -\frac{3}{2} = 5 ( Q -Q_{\rm crit}) 
\ee
For $Q=1/2$ (isotropy) we have $\cA =1$, while with decreasing $Q$ (increasing anisotropy) $\cA$ changes sign and becomes negative;
this happens at $Q =  Q_{\rm crit} = 3/10$. The parameter is minimised at $Q=0$, where $\cA = -3/2$.

Equation (\ref{e1}) describes the time-evolution of the stellar deformation for a given
external tidal field $\cE_{ij}$. In a more compact notation,
\be
\tau_{\nu} \, \dot{e}_{ij} + e_{ij} = \cA^{-1} f_{ij} (t)
\label{e2}
\ee
where, as before, an overdot stands for a time-derivative and a new effective viscous timescale has been
defined:
\be
\tau_{\nu} \equiv \frac{\tau_0}{\cA} 
\label{tvisc}
\ee 
We also note that the source term $f_{ij}(t)$ is identical to that of the isotropic problem. 
Already at this point, and before solving (\ref{e2}), the possibility of $\cA <0$ is an indication
that strongly anisotropic stars would behave in a qualitatively different way under the action of a
given tidal field. This indeed will turn out to be true as we shall shortly see. 

The exact solution of (\ref{e2}) can be easily found (e.g. see Ref.~\cite{poisson09}) but it is not particularly useful or 
physically transparent. A more useful approximate solution can be obtained by assuming that the deformation $e_{ij}(t)$ varies on a timescale $t_f$,
\be
\frac{e_{ij}}{|\dot{e}_{ij}|} \sim t_f \equiv \frac{f_{ij}}{| \dot{f}_{ij} |}
\label{tf}
\ee
Treating $\tau_{\nu}/t_f$ as a small parameter (in agreement with our earlier assumption $\tau_\nu |\Omega-\Omega_{\rm orb}| \ll 1$ 
which led to the approximate Euler equation~(\ref{euler1})) we adopt the following ansatz
\be
e_{ij}(t) = e_{ij}^{(0)}(t) + \frac{\tau_\nu}{t_f} e^{(1)}_{ij} (t) + {\cal O} ( \cA^{-1} f_{ij} \tau^2_\nu/t_f^2 )
\ee
Eqn.~(\ref{e2}) leads to
\be
e_{ij}^{(0)}  = f_{ij}/\cA, \qquad 
e_{ij}^{(1)}  = -t_f \dot{f}_{ij}/\cA
\ee 
which subsequently combine to give
\be
e_{ij}(t) \approx \cA^{-1} \left ( f_{ij}(t) - \tau_\nu \dot{f}_{ij}(t)  \right ) \approx \cA^{-1} f_{ij} (t-\tau_\nu)
\label{esol2}
\ee
This is clearly consistent with (\ref{tf}). Restoring the explicit form of $f_{ij}$,
\be
e_{ij}(t) \approx \frac{5}{4 \cA} \frac{R^3_\star}{M_\star} \left (\, \Omega^2 C_{ij} - \cE_{ij}(t)
+ \tau_{\nu} \dot{\cE}_{ij}(t) \, \right )
\label{esol4}
\ee
This solution looks similar to the classic tidal deformation result: modulo the scaling factor $\cA^{-1}$, the deformation is 
simply equal to the \emph{time-retarded} driving term $f_{ij}$. In a binary system consisting of an orbiting moon and a fluid planet the solution 
(\ref{esol2}) describes a tidal bulge on the planet's surface that follows the position of the moon with a time lag $\tau_\nu$.

This picture is accurate for isotropic and weakly anisotropic stars; increasing the degree of anisotropy ($Q$ moving away from $1/2$ 
towards $Q_{\rm crit} = 3/10$) results in an overall \emph{amplification} of the tidal deformation and a longer time-lag $\tau_\nu$. 
However, the limit  $Q \to Q_{\rm crit}^{+}$ cannot be studied within our formalism because of the divergence in both $\tau_\nu$ and $e_{ij}$.


\subsection{The case of strong anisotropy}

Once the anisotropy parameter $Q$ becomes less than $Q_{\rm crit}$, making $\cA < 0$, the system shows an
unexpected response to tidal perturbations. The exact solution of the deformation equation (\ref{e2}) contains
an exponential term $\sim \exp(t/|\tau_\nu|)$ which becomes dominant at late times $t \gg |\tau_\nu|$, signaling the 
presence of an instability. However, the fact that the characteristic timescale of this solution is $\tau_\nu$ 
casts serious doubts on the physical relevance of this instability: the limit $\nu \to 0$ implies an arbitrarily high
growth rate for the tidal deformation, a prediction that makes little sense.  

It would not be too surprising if  the reader finds the nature of this solution reminiscent of  the `runaway solutions' of the famous Lorentz-Dirac 
equation of motion for an accelerating charged particle (see~\cite{poisson99} for a detailed discussion of this equation). After all, the $\cA < 0$
version of (\ref{e2}) is mathematically equivalent to the Lorentz-Dirac equation. And as it was the case for the unphysical solutions of that equation,
the runaway solution of (\ref{e2}) can be dismissed on the basis of its timescale. It can be easily verified that a solution $e_{ij}/\dot{e}_{ij} \sim \tau_\nu$ 
would generically violate the conditions underpinning the validity of (\ref{euler2}) 
(for instance the inertial acceleration term may outgrow the tidal force) and therefore (\ref{e2}) itself.

The upshot of this discussion is that the tidal deformation of strongly anisotropic stars  ($0< Q < Q_{\rm crit}$) 
should be given by (\ref{esol2}) with $\cA < 0$:
\begin{multline}
e_{ij} (t) \approx -|\cA |^{-1} f_{ij} ( t + |\tau_\nu| ) = 
\\
=-\frac{5}{4 |\cA|} \frac{R^3_\star}{M_\star} \left (\, \Omega^2 C_{ij} - \cE_{ij}(t)
- |\tau_{\nu}| \dot{\cE}_{ij}(t) \, \right )
\label{esol3}
\end{multline}
This solution represents a rather counter-intuitive behaviour for such systems. Firstly, we can see that rotation drives
the body to a \emph{prolate} rather than the usual oblate shape. Secondly, in the presence of a tidal field sourced by an orbiting
satellite we can see that, in the limit of zero viscosity, the induced tidal bulge is at \textit{right angles} to the line pointing 
towards the satellite. In other words the bulge is rotated $90$ degrees with respect to the bulge of isotropic or weakly anisotropic
stars. Viscosity causes an additional slight rotation to the bulge; the bulge's major axis now is orthogonal
with respect to the \emph{time-advanced} position of the orbiting body.



\subsection{Strongly anisotropic systems: `anomalous' quadrupole moment and Love number}

The counter-intuitive tidal dynamics of $\cA < 0$ stars is reflected in the sign of the so-called tidal Love number $k_2$, that is, the proportionality factor appearing in the linear relation between the tidal field and the induced quadrupolar moment, $Q_{ij} = -(2/3) k_2 \cE_{ij}$ where 
\be
Q_{ij} = \int d^3 x\, \rho \left ( x_i x_j -\frac{1}{3} r^2 \delta_{ij} \right )
\label{Qdef}
\ee
is the STF quadrupole moment tensor. This is related to the deformation $e_{ij}$ as~\cite{poisson09}
\be
Q_{ij} = \frac{2}{5} M_\star R^2_\star e_{ij}
\label{Qeq}
\ee

Combing these expressions with the previous solutions for $e_{ij}$, we find, 
\be
k_2 = \frac{3}{4} \cA^{-1} 
\label{love1}
\ee
Normally the Love number is a positive quantity, however, according to our result strongly anisotropic stars 
($\cA <0$) have $k_2 <0$. 

The response of rotating anisotropic stars to the action of the centrifugal force is equally `anomalous' with respect 
to the induced quadrupole moment $\cQ$. This is defined in the usual way, $Q_{ij} = {\rm diag} (- \cQ/3, -\cQ/3,  2\cQ/3)$.
Isolating the centrifugal term in the solution for $e_{ij}$ (i.e. the $C_{ij}$ term) we obtain the quadrupole moment
\be
\cQ = -\frac{R_\star^5 \Omega^2}{2\cA}
\label{Qmom}
\ee
Then, clearly, strongly anisotropic systems have $\cQ>0$ which is of opposite sign with respect to the quadrupole moment
of isotropic stars (this basically means that rotation makes the body prolate rather than oblate).  

These results are of some importance when viewed from the perspective of the "I-Love-Q" universality relations, very
recently shown to exist in neutron and quark stars for a variety of realistic equations of state~\cite{YY13, PA13}.
With increasing stellar compactness, the "I-Love-Q" relations appear to approach the corresponding black hole
parameters, hence suggesting a "no-hair" property for neutron stars~\cite{PA13}. 

Our findings suggest that taking the "black hole limit" of ultracompact stars (i.e.  $R_\star \to 2 M_\star$ ) 
may reveal a \emph{discontinuity} in the "I-Love-Q" relations. Beyond some compactness (that is never reached by neutron 
or quark stars described by any known realistic equation of state), the stellar matter is likely to experience a  "phase-transition" and 
forcibly become strongly anisotropic, thus producing a system with $k_2<0, \cQ >0$ that deviates from a black hole. 
Of course, in order to strictly establish this discontinuity in $k_2,\cQ$ one would have to repeat our analysis 
in full General Relativity and obtain the same sign for these parameters.


\subsection{The tidal torque}

As discussed in Ref.~\cite{poisson09}, the tidal torque exerted on the star is to be calculated in the
global inertial frame (centered at the stellar center of mass). Using an overbar to label quantities
in that frame we have the following expression for the tidal torque:
\be
N_i = -\int  d^3\bar{x} \rho \epsilon_{ijk} \bar{x}^j \partial^k \bar{\Phi}_{\rm tidal}
= -\int d^3\bar{x} \rho \epsilon_{ijk} \bar{x}^j \bar{\cE}^{kl} \bar{x}_l 
\ee
Carrying out the integration we can obtain the rate of change of the stellar spin,
\be
\dot{J}_i = -\epsilon_{ijk} \bar{Q}^j_{~l} \bar{\cE}^{lk} 
\ee
where the quadrupole moment tensor is given in eqns.~(\ref{Qdef}) \& (\ref{Qeq}) above.
Ignoring the possibility of any precessional motion 
(that is, we assume a perturbation that does not change the orientation of $J^i = J \hat{z}^i$ is space) 
we have
\be
\dot{J} = -\epsilon_{ijk} \bar{Q}^i_{~l} \bar{\cE}^{lj} \hat{z}^k
\ee

The coordinate transformation from the rotating to the inertial frame is of course independent of the stellar model.
Hence the entire analysis of Ref.~\cite{poisson09} applies here too without any change. Using the solution (\ref{esol4})
it is found that only the viscous term $\sim \tau_\nu \dot{\cE}_{ij}$ contributes to the tidal torque. It is then a matter of simple 
inspection to conclude that the final result for $\dot{J}$ is that of the isotropic star rescaled with a factor $\cA^{-2}$:
\be
\dot{J} = \frac{1}{\cA^2} \dot{J}_{\rm iso} 
\label{Jdot5}
\ee
It is important to emphasize that a factor $\cA^{-1}$ comes from the rescaled viscous timescale (\ref{tvisc}) while
another similar factor originates from the overall prefactor in the right hand side of (\ref{esol4}). This second contribution
is also present in the \emph{inviscid} system and represents a rescaling in the  tidal Love number
$k_2 \to k_2/\cA$. 

The occurrence of the $\cA^2$ factor means that the tidal torque formula (\ref{Jdot5}) is common for both strongly and weakly
anisotropic stars and, moreover, that $\cA$ and $-\cA$ systems are tidally torqued at the same rate (assuming all other parameters
identical). The fact that $-3/2 \leq \cA \leq 1$ means that anisotropy can enhance the magnitude of the tidal torque, but as the extreme limit
$Q \to 0, \cA \to -3/2$ is approached the effect is reversed.

For the particular case of a tidal field sourced by a test body of mass $\mu$ in a circular equatorial orbit of radius 
$b$ and angular frequency $\Omega_{\rm orb}$ we have
\be
\cE_{ij} = -\frac{3\mu}{b^3} \left ( \hat{m}_i \hat{m}_j - \frac{1}{3} \delta_{ij}  \right )
\ee
where
\be
\hat{m}^i = [ \cos \chi, \sin \chi, 0], \qquad
\chi = (\Omega_{\rm orb}-\Omega)t
\ee
is the unit vector pointing at the instantaneous position of the orbiting body.
After repeating the calculation described in Ref.~\cite{poisson09} the torque (\ref{Jdot1}) becomes
\be
\dot{J} =  \frac{9}{2} \frac{\tau_0}{\cA^2}\frac{\mu^2 R^5_\star}{b^6}  ( \Omega_{\rm orb} - \Omega )
\label{Jdot6}
\ee
This is our final result for the tidal friction on an anisotropic star. Due to the $\cA^{-2}$ factor it displays an 
enhanced torque with respect to isotropic stars, but apart from that it is identical to the classic tidal torque formula.


\section{The superradiance -Newtonian tidal friction correspondence}
\label{sec:newton}

After discussing two incarnations of tidal friction, in the context of black holes (where it was identified with the more exotic
notion of superradiance) and in that of Newtonian fluid stars, we have prepared the ground for making a formal connection between them. 

That there \emph{is} a connection between these two seemingly alien notions had been known for some time~\cite{membrane,hartle73}. 
Apart from repeating this early analysis we make a new contribution to the subject by including anisotropic stars which, as
we have discussed, might be more `realistic' models for ultracompact relativistic bodies.

We have seen that the Newtonian tidal friction torque is given by (eqns.~(\ref{Jdot6}) \& (\ref{tau0}) )
\be
\dot{J}_\rN = \frac{\kappa}{\cA^2}\frac{\nu}{M_\star} \frac{\mu^2 R^6_\star}{b^6} ( \Omega_{\rm orb} - \Omega)
\label{newt1}
\ee
The numerical prefactor $\kappa$ depends on the internal structure of the star, more precisely the
combination $\kappa/\cA$ is a surrogate for the tidal Love number. 
For our uniform star we have $\kappa= 171/4$; given that the Love number for the same system is $k_2 = (3/4) \cA^{-1}$ 
we can write an equivalent expression for the torque by making the substitution
\be
\frac{\kappa}{\cA^2} \to \frac{57 k_2}{\cA}
\ee
The torque (\ref{newt1}) displays the well-known result that a small body 
orbiting rapidly (slowly) with respect to the
central body's rotation will cause the big body to spin-up (spindown) provided viscosity is present.
An equilibrium is reached when the system becomes synchronised, $\Omega_{\rm orb} = \Omega$.

If $\Omega_{\rm orb} \ll \Omega_\star$ (as in Hartle's calculation) the torque (\ref{newt1})
reduces to the spin-down torque:
 \be
\dot{J}_\rN \approx - \frac{\kappa}{\cA^2} \frac{\nu}{M_\star}  \frac{\mu^2 R^6_\star}{b^6} \Omega
\label{newt2}
\ee
In the opposite limit  $\Omega_{\rm orb} \gg \Omega$ (which includes the possibility of a non-rotating star)
we have a spin-up torque:
\be
\dot{J}_\rN \approx \frac{\kappa}{\cA^2}\frac{\nu}{M_\star} \frac{\mu^2 R^6_\star}{b^6} \Omega_{\rm orb}
\label{newt3}
\ee 
These last three Newtonian expressions look very similar to the previous formulae for the torque exerted on a black hole,
see eqns.~(\ref{hartle2}), (\ref{Jdot2}) and (\ref{Jdot4}).

The match between the torque for Newtonian bodies and black holes becomes \emph{exact} if we
make the obvious identifications $R_\star \longleftrightarrow r_{+}$, $\Omega \longleftrightarrow \Omega_\rH$
and subsequently define an \emph{effective} black hole Love number, viscosity and anisotropy:
\be
\frac{k_\rH \nu_\rH}{\cA_\rH} = \frac{M}{570}
\label{effvisc1}
\ee
Then the torque exerted on a black hole, eqn.~(\ref{Jdot4}), becomes
\be
\dot{J}_\rH = \frac{57 k_\rH}{\cA_\rH} \frac{\nu_\rH}{M} \frac{\mu^2 r_{+}^6}{b^6} ( \Omega_{\rm orb} - \Omega_\rH)
\ee

It is clear that the identification (\ref{effvisc1}) implies a high degree of degeneracy between the three effective parameters.
Part of the degeneracy can be lifted if we make use of the effective horizon viscosity appearing in the membrane model of black 
hole event horizons 
\cite{membrane}. The horizon viscosity is given by the remarkably simple formula
\be
\nu_\rH = M  
\label{effvisc2}
\ee
and it allows us to fix the hole's Love number\footnote{This Love number result may seem at odds
with Refs.~\cite{bp09,dn09} which show that the Love number of a Schwarzschild black hole is exactly zero.
We do not believe there is a contradiction here: our Love number is an effective quantity emerging as part of
of the tidal torque identification between a black hole and a material body whereas the vanishing Love number calculated
in \cite{bp09,dn09} is a result of  the computation of the actual event horizon deformation.},
 $k_\rH = \cA_\rH/570$.

Which $\cA$ would be the most natural choice for a black hole? Although $\cA_\rH=1$ could be
a reasonable initial guess (thus making (\ref{effvisc1}) identical to the identification of Ref.~\cite{poisson09}),
a more intuitive choice would be to exploit the $R_\star \to 2M_\star$ property of the Lemaitre `star' and identify 
$\cA_\rH$  with the minimum value of $\cA$, that is,  $\cA_\rH = -3/2$.  In that case the effective Love number
becomes negative, $k_\rH =  -1/380$, a property also shared by the Lemaitre star.

A relation like (\ref{effvisc1}) or (\ref{effvisc2}) is not that absurd; after all, the event horizon is ``viscous'' 
in the sense that it absorbs everything that falls in it. Note that (\ref{effvisc2}) is always used at a level of precision 
at which $r_{+} =  2M + {\cal O}(a^2)$. Hence a relation $\nu_\rH \sim M$ is the only possibility that makes
sense on dimensional grounds (in non-geometric units $M \to GM/c$ in (\ref{effvisc2})). Expressed in human
units, the horizon viscosity is: 
\be
\nu_\rH = \frac{G M}{c} \approx 8.6 \times 10^{14} \left ( \frac{M}{M_\odot} \right )\, \mbox{cm}^2 \mbox{s}^{-1}
\ee
This is an enormous value; it dwarfs by many orders of magnitude the predicted viscosity in neutron star matter --
the most dense state of matter known (see next Section). So for stellar mass compact objects a black hole is far 
more ``viscous,'' for the purposes of tidal friction, than any other compact or hypercompact object is likely to be.
As we shall note, however, for the purposes of comparing to tidal friction in the Earth that an
Earth mass black hole does not have such an enormous viscosity, since the event horizon effective viscosity scales with mass.

The identification between (\ref{Jdot4}) and (\ref{newt1}) via (\ref{effvisc1}) is of key importance for the discussion
of this paper. These formulae (and their limits for large or small ratios $\Omega_{\rm orb}/\Omega$ and $\Omega_{\rm orb}/\Omega_\rH$) 
describe completely different physical mechanisms for producing tidal friction: superradiance as opposed to normal
fluid viscosity. Nevertheless the resulting torques are described by identical formulae, at least within the specified
approximation $ | M (\Omega_{\rm orb} - \Omega_\rH) |, | M (\Omega_{\rm orb} - \Omega) | \ll 1$.

It therefore makes perfect sense to consider superradiance and viscous friction as equivalent notions, in the
sense that they can produce the same torque between the members of a binary system.

In particular, it can be said that a 
Newtonian system can exhibit ``superradiance'', $\dot{J}_\rN < 0$, in the same way a tidally distorted black hole
does. And it is equally legal to consider a black hole as a ``viscous'' body, experiencing tidal friction as
Newtonian fluid bodies do. 

To conclude this discussion it should be pointed out that the tidal friction - superradiance equivalence is unlikely to survive
beyond ${\cal O}(a)$ precision. At higher orders with respect to the spin, the black hole torque is expected to be sensitive to
General Relativistic physics such as frame dragging and ergoregions -- notions completely alien to Newtonian
physics.


\section{Superradiance from horizonless compact objects?}
\label{sec:objects}

The established close connection between relativistic superradiance and Newtonian tidal friction can be exploited if we wish to
understand the gravitational physics of an extreme mass ratio system in which the Kerr hole is replaced by a putative
ultracompact and horizonless massive body. This is not an entirely unreasonable set up: a class of compact horizonless objects --
where gravastars and boson stars are among the most popular members -- is sometimes invoked as a theoretical alternative to
Kerr black holes for explaining the true nature of the supermassive ``dark'' objects commonly found in nuclei of massive galaxies 
(including our own Milky Way).  

Leaving aside the question of how realistic such objects could be (surely none of them has the astrophysical status enjoyed
by black holes and neutron stars) we instead prefer to focus our attention to the notion of ``superradiance'' in these systems.
This is partially motivated by the fact that the simplest compact and horizonless object one could build (albeit in an entirely 
unphysical manner) is a black hole with its horizon enveloped by a perfect mirror.

The resulting ``mirror-Kerr'' object is supposed to mimic 
a horizonless material body in the sense that it is designed 
to reflect, rather than  absorb, 
any incoming gravitational waves. The role of the mirror is to 
provide the total reflection boundary condition, with respect to radial
motion, that the geometric center does in material 
bodies. We ignore the energy removed by the wave and 
dissipated as heat as a result of its coupling with the body's matter.
Note, however, that doing so is probably tantamount to
presuming that the body is inviscid!

The presence of the mirror has a striking effect on superradiance. In fact, superradiance disappears altogether.
This was recently pointed out in Ref.~\cite{richartzetal} in the context of scalar wave propagation in Kerr spacetime. 
The same conclusion remains true in the case of gravitational waves -- this is discussed in detail in a later Section. 

Extrapolation of this result to the case of more realistic horizonless bodies would imply that these objects
do not display any superradiance of the General Relativistic type. This certainly can be a correct statement, however,
it does not account for the likely presence of Newtonian-type ``superradiance'' in the form of tidal friction. 
As we have discussed, at leading order the two effects are dynamically equivalent. What is really required of the horizonless
object is to consist of viscous matter.  

To make the discussion more concrete let us consider the particular example of the anisotropic `star' of Section~\ref{sec:torque}.
Some degree of anisotropy in the fluid pressure seems to be a key ingredient in relativistic stellar models with high compactness,
such as gravastars (these objects consist of a de Sitter spacetime vacuum spherical interior which is cloaked by a thin shell of high-density 
anisotropic fluid matter, see~\cite{cattoen05}). 

Our earlier result for the tidal torque exerted on an anisotropic star by an orbiting test-body, eqn.~(\ref{newt1}),
can be translated to an effective ``horizon'' energy flux (this is identical
to the tidal work $\dot{W}$ of Ref.~\cite{poisson09}):
\be
\dot{E}_\star = \Omega_{\rm orb} \dot{J}_\rN
= \frac{\kappa}{\cA^2} \frac{\nu}{M_\star} \frac{\mu^2 R_\star^5}{b^6} \Omega_{\rm orb} ( \Omega_{\rm orb} - \Omega )
\label{Estar1}
\ee
For orbital motion with  $\Omega_{\rm orb} < \Omega$ the central object exhibits ``superradiance'', $\dot{E}_\star < 0$, and this flux is gained 
by the orbiting body itself as a result of angular momentum conservation. 

At the same time the small body radiates  gravitational waves at infinity with a flux (given here at a quadrupole-formula precision)
\be
\dot{E}_\infty \approx \frac{32}{5} \frac{\mu^2}{M^2_\star} v^{10} 
\ee

For obtaining a numerical value for the flux (\ref{Estar1}) we obviously need to know the coefficient of kinematic
viscosity $\nu$. Unfortunately this is not possible: the detailed composition -- let alone properties -- of matter
at such high densities as the ones required to build a supermassive-ultracompact star with $R_\star \approx 2M_\star$ is completely
unknown (assuming that such objects are realised in Nature). 
Nonetheless we can get some idea of how viscous dense matter is by making contact with neutron stars --
the most dense astrophysical objects known to us\footnote{The next densest astrophysical objects, white dwarfs, have a 
much smaller shear viscosity, see for instance~\cite{shternin08}.}. A typical viscosity for neutron star matter is~\cite{CL87},
\be
\nu_{\rm ns} \approx 10^{4}\, \left ( \frac{\rho}{10^{14}\,\mbox{gr}\, \mbox{cm}^{-3}} \right )^{5/4} 
\left ( \frac{10^8\,\mbox{K}}{T} \right )^2\, \mbox{cm}^2 \mbox{s}^{-1}
\label{NSvisc} 
\ee
where $T$ is the neutron star's core temperature and $\rho$ its density (both quantities have been normalised
to values typical for mature neutron stars).

The viscosity (\ref{NSvisc}) could be, perhaps, viewed as a lower limit for the viscosity in supermassive ultracompact stars. 
We can also come up with an upper limit using the general kinetic theory formula for the coefficient of dynamic and kinematic 
shear viscosity, $\eta$ and $\nu$,
\be
\nu = \frac{\eta}{\rho} =  \frac{1}{5} \, c_s \lambda 
\label{kinetic1}
\ee
Here $c_s$ is the speed of sound, $\lambda$ is the mean free path length associated with the particle collisions responsible
for viscosity and $\rho$ is the density of these particles. 

From this relation we can obtain a maximum viscosity by setting $c_s = c$  
and $\lambda = 2R_\star$, that is (we temporarily restore $c$)
\be
\nu_{\rm max}  = \frac{2}{5} c R_\star
\label{kinetic2}
\ee
For neutron star matter we have $c_s \sim 0.1 c$ and $\lambda \ll R_\star$ and as a result $\nu_{\rm ns} \ll \nu_{\rm max}$.
The effective black hole horizon viscosity $\nu_\rH$ does not really have to do anything with particle collisions but nevertheless
we can still compare it against (\ref{kinetic2}) with $R_\star$ replaced by $r_{+} = 2GM/c^2$. Using our previous result we find
\be
\frac{\nu_{\rm ns}}{\nu_{\rm max} (R_\star)} \ll \frac{\nu_\rH}{\nu_{\rm max} (r_{+})}  \sim 1
\ee
Most notably, the effective horizon viscosity is comparable to the maximum viscosity $\nu_{\rm max}$.

The upshot of this discussion is that a `realistic' ultracompact star could be, at least in principle, as viscous as
a black hole or thereabout. Combined with the enhancement due to the anisotropy factor $\cA < 1$ this would imply that the flux $\dot{E}_\star$
\emph{might} be an important factor in the gravitational wave-driven inspiral of small bodies around such stars. If nothing else, these
systems can be ``superradiant'' even if negligibly so. 

Continuing along the same line of reasoning we could, for example, ask if ``floating'' orbits can exist. 
By definition these are test-body orbits where the orbital energy lost to gravitational radiation at infinity is
balanced by the horizon flux of the central object. It is known that such orbits are not possible around black holes (see~Ref.~\cite{floatpaper}
for a very recent discussion) but this may not be necessarily true for other compact systems. 
For the present case such orbits would obey $\dot{E}_\infty = -\dot{E}_\star$. The resulting orbital radius is
\be
\frac{b_f}{M} \approx \left ( \frac{855}{128} \frac{\Omega}{\Omega_\rK} \right )^{2/5} 
\cA^{-4/5} \left ( \frac{\nu}{M_\star} \right )^{2/5} \left ( \frac{R_\star}{M} \right )^{9/5}
\ee 
Note that this expression was derived using a weak-field $b_f \gg M$ approximation but it could also be indicative of
strong field orbits. As implied by our previous discussion, a prerequisite for an orbit to float is the presence of  
highly viscous matter $\nu \sim \nu_{\rm max}$, assisted by an anisotropy factor $|\cA| < 1$.
In an earlier paper \cite{floatpaper} we showed that for a floating orbit
to exist in a compact binary system, one needs the central spinning body to 
have a prolate shape. It is worth recalling
that anisotropic pressure naturally produces a prolate shape in
a rotating body (see the discussion after eqn.~(\ref{esol3})).

As we have argued, there appears to be strong reasons to believe that ultracompact
relativistic stars with viscous interiors do experience a phenomenon
which can be equivalently described as tidal friction or superradiance,
rather as is the case for black holes. In the next section we will
show that eliminating viscosity, which would clearly reduce tidal 
friction to zero, also eliminates superradiance. 

The question naturally arises whether ordinary material systems (such
as the Earth-Moon) also can be thought of as superradiant, when they
are observed to undergo tidal coupling. It is not the purpose of our paper to
consider this issue, but it is worth noting that if one replaced the Earth
with a black hole of the same mass, the effective viscosity of that black
hole would be not too dissimilar to that of the present
Earth. Thus, an Earth-sized black hole would not be appreciably better
at absorbing gravitational waves than the Earth is. While of course the
Earth-sized black hole would have far smaller tidal friction than the
Earth does, we would still expect the tidal friction it did exhibit to
be describable in terms of gravitational wave superradiance. 


\section{Superradiance in the presence of a mirror}
\label{sec:mirror}

The comparison between black holes and ultracompact relativistic stars can also take a different form, one in which the `star' is modelled
as a black hole enveloped by a `mirror' surface. Clearly this `mirror-Kerr' body is a purely theoretical construct which, nevertheless, 
possesses some of the properties of more realistic relativistic stellar models. 

From a mathematical point of view the role of the mirror is to mimic the total reflection imposed on the radial motion of waves at the
stellar center. Moreover, the effective potential governing wave propagation
is qualitatively similar to that of compact stars (e.g. the system is endowed with 
spacetime $w$-modes~\cite{KKreview}). As far as superradiance is concerned it is known that the addition of the mirror eliminates it altogether. This
has been recently pointed out, for scalar waves, by Richartz et al.~\cite{richartzetal}. 

Our discussion in this Section considers the interaction of gravitational waves with a mirror-Kerr object. To this end
we first generalise the analysis of~\cite{richartzetal} and reach the same conclusion as they did, namely, the absence of
superradiance. However, we also take few more steps and make contact with the so-called ergoregion instability. This phenomenon
is characteristic of ultracompact stars and is also a property of the mirror-Kerr system~\cite{vilenkin78}. In the end, we
provide a unified description of the transition from superradiance to the ergoregion instability.

       
\subsection{Superradiant scattering of gravitational waves}

Studying the scattering of waves (gravitational or not) by a Kerr black hole is greatly facilitated by the well-known Teukolsky 
formalism~\cite{teuk73},\cite{tp74}. The required wave equation is not the original Teukolsky equation but its Schr\"odinger-type variant.
In the notation of~\cite{tp74} (their eqn.~(5.2)) this equation is
\begin{equation}
\frac{d^2Y}{dr_{*}^2} + V Y = 0
\label{YTeuk}
\end{equation}
where $Y$ is the radial wavefunction, $r_*$ is the usual tortoise coordinate and $V$ plays the role of the scattering potential 
(its specific form will not be needed here).

As detailed in~\cite{tp74} the constant Wroskian property of (\ref{YTeuk}) can be applied to the two independent solutions 
$Y(s)$ and $\bar{Y}(-s)$, where $s$ is the spin-weight of the field in question ($s=\pm 2$ for gravitational perturbations) 
and an overbar denotes a complex conjugate. Equating the Wronskian evaluated at infinity and at the horizon leads to a relation
between the wave amplitudes at these two boundaries. This relation is the one predicting wave superradiance when the 
frequency obeys $ \tilde{\omega} = \omega - m \Omega_\rH < 0$ (here $m$ is the usual spherical harmonic integer mode associated
with the azimuthal coordinate $\varphi$). 

For our own discussion of gravitational wave scattering we will follow a slightly different path, namely, one that makes more contact
with the scalar wave analysis of~\cite{richartzetal} and that can be phrased in terms of reflection and transmission amplitudes. 
For this purpose, the wave equation (\ref{YTeuk}) is not a suitable equation because its solutions are not pure plane waves (for example,
at infinity we have solutions of the form $Y(r\to \infty) \sim r^{\pm s} e^{\mp i\omega r_*} $). This unwelcome property is a direct consequence
of the long-range character, $\sim 1/r$, of the $V$ potential.

A more suitable equation can be build with the help of the Sasaki-Nakamura formalism (see Ref.~\cite{chapter} 
for a review), one which by design features a short-ranged potential. Written in a Schr\"odinger form and for a fixed 
spin $s=-2$ this equation is (see~\cite{gk02} for details):
\begin{equation}
\frac{d^{2}X}{dr^2_*} + V_{\rm SN} X = 0
\label{SNeq}
\end{equation}
where $X$ is the new wavefunction.
The Sasaki-Nakamura potential $V_{\rm SN}$ has the asymptotic behaviour  
\bear
&&V_{\mathrm{SN}}(r \rightarrow \infty) = \omega^2 + \cO(1/r^2)
\\
&& V_{\mathrm{SN}}(r \rightarrow r_{+}) = \tilde{\omega}^2 + \cO\left(e^{(r_{+} - r_{-})/2M}\right) 
\eear
and indeed it decays faster at radial infinity than the Teukolsky potential $V$.

Eqn.~(\ref{SNeq}) admits plane wave solutions at infinity and at the hole horizon. In particular the solution $X_{+}$, describing incoming waves
(amplitude $A_{\rm in}$) which are partially reflected (amplitude $A_{\rm out}$) and partially transmitted down the hole
(amplitude $A_{\rm tr}$), has the asymptotic form
\begin{eqnarray}\label{xplus}
&& X_{+}(r \rightarrow \infty) = A_{\mathrm{in}}e^{-i\omega r_{\star}} + A_{\mathrm{out}}e^{i\omega r_{\star}} 
\\
&& X_{+} (r \rightarrow r_{+}) = A_{\mathrm{tr}}e^{-i\tilde{\omega}r_{\star}}
\end{eqnarray}
We can then define reflection and transmission coefficients in a standard way,
\be
\cR = \frac{A_{\rm out}}{A_{\rm in}},\quad \cT = \frac{A_{\rm tr}}{A_{\rm in}}
\ee
The $A$ amplitudes are related to the amplitudes of the solutions of the Teukolsky equation (\ref{YTeuk}) in a simple way~\cite{gk02} 
(these are the $B$ amplitudes in Ref.~\cite{tp74}), thus permitting the connection of $\cR$ and $\cT$ using the Wronskian equality $W(r\to \infty) = W (r=r_{+})$ of that latter equation. 
The result of this exercise is
\begin{equation}
|\mathcal{R}|^2 = \frac{|c_{0}|^2}{|\mathcal{C}|^2}\left (\, 1-\frac{\tilde{\omega}}{\omega}\mathcal{D}|\mathcal{T}|^2 \, \right )
\label{reflection}
\end{equation}
where $c_0$, $\mathcal{C}$ (the so-called Starobinski constant)  are complex functions of the wave-frequency $\omega$ and the black hole
parameters $M,a$. Although their detailed forms are not needed, we note that they can be found in Ref.~\cite{chapter}.
The parameter $\cD$ is real and positive and is given by:
\begin{equation}
\mathcal{D}(\omega, \tilde{\omega}) = \frac{16(2Mr_{+})^5(\tilde{\omega}^2 + 4\epsilon^2)(\tilde{\omega}^2 + 16\epsilon^2)
|c_{0}|}{|\eta||d|^2}
\end{equation}
with
\begin{eqnarray}
d &=& \sqrt{2Mr_{+}}\left[8M^2 - 4a^2m^2 - 12iamM \right. \\
  &+& \left. (8-24i\omega M - 16\omega^2 M^2) \right. \nonumber \\           
  &+& \left. (-16M + 16am\omega M + 12iam + 24i\omega M^2)r_{+}\right] \nonumber
\end{eqnarray}
The functions $\eta$ and $\epsilon$ can also be found in the Sasaki-Nakamura formalism section of Ref.~\cite{chapter}.

At first glance the result~(\ref{reflection}) seems counterintuitive: in some part of the parameter space $|c_{0}/\cC| > 1$, implying a reflection
$|\cR| > 1$ and thus suggesting the existence of superradiance even when $\tilde{\omega} > 0$, contrary to established results \cite{tp74}. 
However, superradiant amplification must be formulated in terms of the ratio of the incoming energy flux to that of the outgoing radiation.
Unlike the case of scalar waves where this ratio is identical to the amplitude ratio, i.e. 
$\cR = \dot{E}_{\rm out}/\dot{E}_{\rm in}  = |A_{\rm out} |^2 / | A_{\rm in} |^2$, the gravitational wave fluxes are related to the amplitudes in 
a more subtle way. Expressing the gravitational wave fluxes from~\cite{tp74} in terms of the Sasaki-Nakamura amplitudes,
\begin{eqnarray}
&\dot{E}_{\mathrm{in}}& = \frac{8\omega^2}{|\cC|^2|c_0|}|A_{\mathrm{in}}|^2 
\label{energ1}
\\
&\dot{E}_{\mathrm{out}}& = \frac{8\omega^2}{|c_0|^3}|A_{\mathrm{out}}|^2 
\label{energ2}
\\
&\dot{E}_{\mathrm{tr}}& = \frac{8\omega\tilde{\omega}\cD}{|\cC|^2|c_0|}|A_{\mathrm{tr}}|^2
\end{eqnarray} 
With these we can define a new pair of reflection and transmission coefficients:  
\begin{eqnarray}
&\cR_{\mathrm{flux}}& = \frac{\dot{E}_{\mathrm{out}}}{\dot{E}_{\mathrm{in}}} = \frac{|\cC|^2}{|c_{0}|^2}|\mathcal{R}|^2 \\
&\cT_{\mathrm{flux}}& = \frac{\dot{E}_{\mathrm{tr}}}{\dot{E}_{\mathrm{in}}} = \frac{\tilde{\omega}}{\omega}\cD|\cT|^2  
\end{eqnarray}
Then using \eqref{reflection}
\begin{eqnarray}
\cR_{\mathrm{flux}} = 1-\frac{\tilde{\omega}}{\omega}\mathcal{D}|\mathcal{T}|^2 = 1-\cT_{\mathrm{flux}}
\label{infinityRatio}
\end{eqnarray}
This expression re-establishes the standard condition for superradiance: $\tilde{\omega} < 0 \Leftrightarrow \cT_{\mathrm{flux}} < 1$ 
and $\cR_{\mathrm{flux}} > 1$.

At this point, if we replace the horizon with a perfect mirror we should have $A_{\mathrm{tr}} = 0 \to \cT = \cT_{\rm flux} = 0$.
In other words, the presence of the mirror fixes $\cR_{\mathrm{flux}} = 1$ and eliminates superradiance altogether 
regardless of $\tilde{\omega}$. This is in agreement with the findings of~\cite{richartzetal} for the black hole-scalar waves system. 


\subsection{Wave and Particle Superradiance}

Having argued that black hole tidal friction and superradiant scattering are just two ways of looking at the same unique 
wave-amplification mechanism, it is worthwhile to ask whether another mechanism which extracts energy and angular
momentum from the black holes, the Penrose process, is in some way related. Descriptions of superradiant scattering as simply
the Penrose process for waves (instead of particles) are fairly common in the literature
and even more so in classroom accounts. However there is a difficulty with this comparison, which is that the Penrose process can take place
only with the ergoregion of a black hole (or other sufficiently hypercompact body) but superradiance does not depend upon the ergoregion
for its existence. This has recently been pointed out by Richartz \& Saa \cite{RichartzSaa13}, who observe that the first discussion of 
superradiance, by Zel'dovich~\cite{Zel'Dovich71,Zel'Dovich72}, involved a rotating cylinder with no ergoregion. While it is true that 
superradiance requires negative energy modes in the system with which the scattered wave interacts, this can be achieved without
an ergosphere.

The Penrose process relies crucially on the existence of the ergosphere, wherein geodesics with negative energies exist. This allows for particle decay processes occurring inside the ergosphere to transfer a portion of the black hole's energy and angular momentum to decay products that exit the ergosphere.

For slowly rotating black-holes, the radius of the ergosphere, when expanded to first order in $a$, coincides with the event horizon radius: 
\begin{equation}
r_{\mathrm{ergo}} = r_{+} = 2M + \cO(a^2)
\end{equation}
There is therefore $no$ ergoregion at $\cO(a)$, making it impossible for the Penrose process to occur at this order.

On the other hand, expanding the energy flux reflection coefficient (cf. \eqref{infinityRatio}) to first order in $a$, we get:
\begin{equation}
\cR_{\mathrm{flux}} = 1-q_0|\cT |^2 + \left(\frac{mq_0}{4M\omega} - q\right)|\cT |^2 a + \cO (a^2)
\end{equation}
where the constant $\cD$ in \eqref{infinityRatio} is expanded as $\cD = q_0+qa + \cO(a^2)$ ($q_0$ and $q$ are constants that 
are independent of $a$ but, in general, are complicated functions of the orbital radius $r$ and harmonic integer $m$). 

For sufficiently high values of $m$ and a range of radii $r$, it can be shown that superradiance ($\cR_{\mathrm{flux}} > 1$) could persist at $\cO (a)$. Contrast this with the Penrose process which is completely dependent on the ergoregion, a second order effect in $a$. Therefore, even though both mechanisms are forms of scattering that steal the black hole's energy and angular momentum, they are fundamentally distinct processes.



\subsection{From superradiance to the ergoregion instability}
\label{sec:ergo}

The disappearance of superradiance is not the only change caused by the replacement of the event horizon with a mirror.
A perhaps more dramatic consequence is the appearance of the ergoregion instability. This is a dynamical type of 
instability, already known in the context of compact relativistic stars~\cite{friedman78,comins78}. 
If sufficiently compact and rapidly rotating, such stars have an ergoregion and a potential `cavity' between the stellar center and the peak of the potential in the vacuum exterior. 
The instability sets in through the system's trapped $w$-modes~\cite{KRA04}; 
these are the spacetime modes associated with the cavity of the potential.

Our system consisting of waves propagating in a mirror-Kerr field is naturally prone to the ergoregion instability. In fact, this is the
same system considered by Vilenkin~\cite{vilenkin78} (albeit for scalar waves) in one of the first discussions of the ergoregion
instability, and much before it was associated with unstable $w$-modes. The required cavity in the potential ($V$ or $V_{\rm SN}$) is formed 
between the mirror and the peak of the potential (the peak is located near the Kerr spacetime's unstable circular photon orbit). This is also
where the hole's ergoregion is located\footnote{The set up described here bears a strong similarity with the so-called `black hole bomb' mechanism
devised by Press \& Teukolsky~\cite{PT72} where a cavity is formed between the event horizon and a mirror placed outside the
potential peak.}. 

Impinging radiation with frequency $\omega$ can become unstable through the ergoregion mechanism provided $\omega < m\Omega_{\rm H} $,
i.e. the same condition that would lead to superradiance if the mirror was not there. In this sense superradiance and the
ergoregion instability seem to be two sides of the same coin. However, and unlike the puny amplification caused by superradiance,
the ergoregion instability leads to an exponentially growing wave reflected back to infinity. In the language of normal modes,
the instability is associated with a complex-valued eigenfrequency. This is also the reason why this effect was not present in 
the previous real-frequency analysis. 

How the exponential growth comes about can be very intuitively understood by the scattering experiment described
in the Vilenkin paper~\cite{vilenkin78}. This consists of an initially incoming pulse, partially reflected and partially
transmitted through the black hole potential. The transmitted part gets reflected at the mirror and becomes radially outgoing.
Then this secondary wavepacket gets partially reflected and partially transmitted. If the condition $\tilde{\omega} < 0$ is satisfied the 
transmitted part escapes to radial infinity after having been amplified in the ergoregion. The same process is repeated
everytime a sub-pulse is reflected towards the black hole. As a result of this an observer far away registers an exponentially 
growing wave signal.

This procedure is essentially the same for scalar and gravitational waves and therefore it would be sufficient to sketch 
the calculation here and refer the reader to Ref.~\cite{vilenkin78} for full details. However, apart from considering gravitational perturbations, 
we expand the original 
calculation in one more way: we accommodate the possibility of
an \emph{imperfect} mirror, meaning that the mirror causes only partial reflection while absorbing the rest of the impinging 
wave. Thus, the imperfect mirror allows a smooth transition from an event horizon (total absorption)
to a perfect mirror (total reflection) type of boundary condition. Technically this would  mean that an incoming wave of amplitude 
$D_{\rm inc}$ just before interacting with the mirror is reflected back with an amplitude
\be
D_{\rm ref} = \lambda D_{\rm inc}, \qquad 0 \leq \lambda \leq 1
\label{mirror}
\ee   
Then $\lambda=1$ would correspond to Vilenkin's perfect mirror system while $\lambda=0$ can be thought as the event horizon limit.

The ergoregion instability calculation is based on the use of two linearly independent solutions of (\ref{SNeq}). The first solution
is the one given earlier, $X_{+}$, with the incoming amplitude set to unity, $A_{\rm in}=1$. The second solution, denoted as $X_{-}$, 
has an asymptotic behaviour
\begin{eqnarray}
&& X_{-}(r \rightarrow \infty) = A_{\mathrm{up}}e^{i\omega r_{\star}} 
\nonumber \\
&& X_{-} (r \rightarrow r_{+}) = A_{\mathrm{down}}e^{-i\tilde{\omega}r_{\star}} +  e^{i\tilde{\omega} r_{\star}}
\label{xminus} 
\end{eqnarray}
This solution represents a plane wave of unit amplitude moving radially outwards from the vicinity of the horizon/mirror
and being partially transmitted to infinity (amplitude $A_{\mathrm{up}}$) and partially reflected back 
(amplitude $A_{\rm down}$). 

The first relation between wave amplitudes corresponding to the $X_{+}$ solutions (cf. eqn.~\eqref{xplus}) is in fact nothing but 
eqn.~\eqref{reflection} with unit incident-wave amplitude $A_{\mathrm{in}} = 1$ : 
\begin{equation}
|A_{\mathrm{out}}|^2 = \frac{|c_{0}|^2}{|\mathcal{C}|^2} \left ( 1-\frac{\tilde{\omega}}{\omega}\mathcal{D}|A_{\mathrm{tr}}|^2 
\right )
\label{wro1}
\end{equation}
The second set of relations corresponding to the $X_{-}$ solutions may be trivially found by comparing eqns. \eqref{xplus} and \eqref{xminus}. 
A variable transformation of the form $r_{\star} \rightarrow -r_{\star}$ leaves the Sasaki-Nakamura equation \eqref{SNeq} functionally unchanged. 
Therefore, the $X_{+}$ and $X_{-}$ solutions are symmetric under this transformation, provided we swap the variables $\omega \leftrightarrow \tilde{\omega}$. 
Interchanging these variables in eqn.~\eqref{wro1} (and appropriately labeling the amplitudes), the amplitude relation 
corresponding to the $X_{-}$ solutions is found to be:
\begin{equation}
|A_{\mathrm{down}}|^2 = \frac{|\tilde{c}_{0}|^2}{|\mathcal{\tilde{C}}|^2}
\left (1 -\frac{\omega}{\tilde{\omega}}\mathcal{\tilde{D}}| A_{\mathrm{up}}|^2 \right )
\label{wro2}
\end{equation}
where $\tilde{c}_0 = c_0 (\tilde{\omega})$, $\tilde{\cC} = \cC(\tilde{\omega})$ and $\tilde{\cD} = \cD(\tilde{\omega}, \omega)$.

The total gravitational wave energy $E$ registered at infinity with respect to the energy $E_0$ of the initial pulse can be
calculated with the help of these amplitude relations, the reflection condition (\ref{mirror}) and the energy formula (\ref{energ1}).
Essentially, this is the same quantity as the reflection coefficient, i.e. $\cR_{\rm flux}= E/E_0$. The result is: 
\be
\cR_{\rm flux} =  1-\frac{\tilde{\omega}}{\omega}\mathcal{D}|A_{\mathrm{tr}}|^2 
+ \lambda^2\tilde{\cD}\cD |A_{\mathrm{tr}} A_{\mathrm{up}}|^2
\sum_{n = 0}^{\infty}\left(\frac{\lambda}{\lambda_0} \right)^{2n}
\label{Eratio0}
\ee
with
\be
\lambda_0 \equiv \frac{|\tilde{c}_0|}{|\tilde{C}|} \frac{1}{|A_{\mathrm{down}}|}
\ee
This result generalises Vilenkin's formula (his eqn.~(5)) to the case of gravitational waves and an imperfect mirror.

Consider first the case $\tilde{\omega} > 0$. From \eqref{wro2} this implies $|\tilde{C}|^2 \lambda^2|A_{\mathrm{down}}|^2/ |\tilde{c}_0|^2| < 1$ 
for the range of allowed values of $\lambda$. Hence the infinite summation in \eqref{energ1} converges, and we have:
\be 
\cR_{\rm flux} = 1-\frac{|\tilde{\omega}|}{\omega}\mathcal{D}|A_{\mathrm{tr}}|^2 + \lambda^2\tilde{\cD}\cD\frac{ |A_{\mathrm{tr}} A_{\mathrm{up}}|^2}{1-\lambda^2/\lambda_0^2}
\label{Eratio1}
\ee  
It is easy to verify that for $\lambda=1$ (the perfect mirror limit) this expression reduces to $\cR_{\rm flux} =1$, in other words the 
entire energy of the initial pulse is reflected back to infinity. When $\lambda < 1$ part of the initial energy is absorbed 
by the hole, naturally resulting in $\cR_{\rm flux} < 1$ (note that $\tilde{\omega}>0$ implies $\lambda_0 >1$, see eqn.~(\ref{wro2})).

The phenomenology is much more interesting when $\tilde{\omega} < 0 $. This entails $\lambda_0 < 1$ and then we need to consider separately 
the subcases $\lambda_0 < \lambda \leq 1$ and $\lambda < \lambda_0$. To begin with, we first note that $\lambda=0$ (the event horizon limit)
leads to
\be
\cR_{\rm flux} = 1 + \frac{|\tilde{\omega}|}{\omega}\mathcal{D}|A_{\mathrm{tr}}|^2 > 1
\label{Eratio2}
\ee 
This is the standard superradiance result.

For $\lambda < \lambda_0$ we again obtain $\cR_{\rm flux} > 1$ but this time the superradiance is \emph{enhanced} with respect to
the previous result:
\begin{equation}
\cR_{\rm flux} = \cR_{\rm flux} |_{\lambda=0} + \lambda^2\tilde{\cD}\cD\frac{ |A_{\mathrm{tr}} A_{\mathrm{up}}|^2}{1-\lambda^2/\lambda_0^2}
\label{Eratio3}
\end{equation}

Finally, the ergoregion instability emerges when $\tilde{\omega} < 0$ and for a mirror reflectivity $\lambda_0  < \lambda \leq 1$. 
Then the sum in (\ref{Eratio0}) is divergent, making $\cR_{\rm flux} \to + \infty$ (note that this divergence is a prediction of linear perturbation
theory -- once the wave amplitude has grown sufficiently, non-linear effects would become important and should be accounted for).
Hence, the critical value $\lambda=\lambda_0 (\omega)$ marks the transition from a monotonically increasing superradiance to the ergoregion instability. 

The fact that an imperfect mirror restores superradiance, enhances it and eventually promotes it to the ergoregion instability can now
be viewed from the perspective of a viscous material body. The imperfect mirror, with $\lambda < 1$ is 
a dissipative surface because we associate it with a material body with the property that it can
absorb a portion of the gravitational radiation which passes through it. Where a material body has a
very high viscosity, and therefore can absorb a significant proportion of gravitational wave energy,
then the mirror becomes significantly imperfect and superradiance is restored.


\section{Concluding discussion}
\label{sec:conclusions}

When comparing the superradiant scattering which occurs in a binary black hole system with the tidal friction
in a material system like our Earth and its Moon, one can initially be struck by the indirectness of the analogy
proposed by Hartle~\cite{hartle73,hartle74}. This analogy, that superradiant scattering of gravitational waves in
the black hole binary be treated just as if it was a case of tidal friction between two orbiting bodies, was 
subsequently elaborated further in the membrane paradigm~\cite{membrane}.

Tidal friction, as its name implies, relies on some form of dissipation to facilitate the angular momentum transfer
between the two bodies in the system. Since it is impossible, in the unequal mass case, to satisfy both
energy and angular momentum conservation, some dissipation must occur. 
In the case of material bodies, this dissipation is due to tidal heating. In the black hole case the dissipation is due to the event horizon's ability to perfectly absorb gravitational radiation. Clearly material bodies are not good absorbers of gravitational energy.
They are virtually transparent to gravitational waves.

For this reason, when replacing the black hole with a material body, a typical model is to think of it as a black hole with a mirror placed
just in front of the horizon. The notion is that the material body is virtually transparent to gravitational waves, 
which pass right through and (speaking radially), come right back out again, as if reflected. It has been shown by 
Richartz et al. \cite{richartzetal}, and confirmed earlier in this paper, that such a mirror destroys superradiance. 
Yet tidal friction is certainly well known in material bodies, so why do we persist in thinking that superradiance and 
tidal friction are equivalent, if destroying one leaves the other intact? 

To answer this question it is instructive to consider replacing like with like. The effective viscosity of the black hole
event horizon (for black holes of stellar mass or greater) is a huge quantity, many orders of magnitude higher than even the 
viscosity of bodies like the Earth (which is high by the standards of most materials). What would happen if we replaced the black 
hole with a similarly sized (in mass and radius) physical body with a viscosity close to that of the event horizon? It seems clear
that the mirror model would not be an appropriate one in that case. Most materials are nearly transparent to gravitational waves, 
but hypothetical materials with such large viscosity would not be. The absorption of gravitational waves by a viscous fluid has
been studied in the weak field case by Esposito \cite{Esposito71} and in the strong field case (of a fluid ball close to a Schwarzschild
black hole) by Papadopoulos \& Esposito \cite{PapadopoulosEsposito85}. They show clearly that an arbitrarily viscous fluid
will be arbitrarily efficient at absorbing incident gravitational radiation. Since this is the case, it strongly implies that 
a hyper-compact highly viscous fluid star with an orbiting satellite would exhibit tidal friction resulting in a transfer of the 
star's angular momentum to the satellite, since the dissipation would consist of the near-complete absorption of the incident gravitational 
energy from the orbiting satellite.

The parallel between tidal friction and superradiance has rarely been mentioned in print in recent years, though an important exception
is provided by the interesting paper of Cardoso \& Pani~\cite{cardpan}. One question which naturally arises, which is also discussed by them,
is whether the case of dissipation in material bodies, mediated through viscosity, is characteristically different from tidal friction in the black hole case,
in which the dissipation is provided by the absorptive capacity of the event horizon. It might be tempting to interpret the viscosity mediated kind of 
tidal friction in material bodies as an electromagnetic mediated kind of interaction, on the grounds that viscosity is an electromagnetic process.
It seems to us, however, that the electromagnetic tidal friction is of the type discussed in Zel'dovich~\cite{Zel'Dovich71,Zel'Dovich72}, and 
Richartz \& Saa~\cite{RichartzSaa13}, in which the electromagnetic field actually mediates between the two bodies, and in which one can think of the
effect as electromagnetic superradiance. Accordingly one should think of the black hole and the hypercompact ultraviscous star as fully
equivalent systems.

One other possible source of absorption, without imagining hyperviscous materials, is the possibility of a hypercompact object trapping
gravitational waves inside its own potential barrier, producing what are known as $w$-modes, gravitational waves trapped close to a
hypercompact object (more compact than a neutron star, for instance). An orbiting particle around such a body would send down gravitational waves which would temporarily disappear behind the potential barrier, creating $w$-modes, and transferring angular momentum to the orbiting body. If the $w$-modes are long lived, then the particle could spiral in rapidly and plunge down to merger
before it ever gets back the angular momentum. 

In our investigation of the superradiance-tidal friction analogy we have taken a closer look to the tidal interaction
between a fluid star made of anisotropic and viscous matter and an orbiting small body. The choice of anisotropic matter
was dictated by the need of modelling a `star' with such a high degree of compactness that could be ``mistaken'' for
a (supermassive) black hole by observations. Although our analysis was Newtonian, and therefore not strictly applicable 
to ultracompact relativistic objects, we have found an interesting set of results. Systems with weakly anisotropic matter 
show a response to tidal fields similar to that of isotropic stars. On the other hand, strong anisotropy causes a
rather counter-intuitive response, with respect to both rotational and tidal forces: rotation makes the shape of the body prolate 
rather than oblate, the tidal Love number becomes negative and the tidal bulge is rotated by roughly 90 degrees with respect to its orientation in isotropic of weakly anisotropic stars.
At the same time, and regardless of the degree of anisotropy, the resulting tidal torque is given by a single universal formula. 
These results are clearly interesting and deserve further and more detailed work using the full machinery of relativistic
gravity and/or the use of the full Euler equations. We hope to be able to report more progress on this subject in the near future.


\section*{Acknowledgements}
KG is supported by the Ram\'{o}n y Cajal Programme of the Spanish Ministerio de Ciencia e Innovaci\'{o}n 
and by the German Science Foundation (DFG) via SFB/TR7. The authors would like to thank Stephen O'Sullivan
and Eric Poisson for fruitful discussions.  

\appendix

\section{The linearised Euler Equation}
In this Appendix we give an outline of the derivation of the linearised Euler equation \eqref{fulleuler} describing 
the perturbed state of an anisotropic fluid in Newtonian gravity. This is done by first deriving the relevant 
hydrodynamical equations in General Relativity and then taking their Newtonian limit.

The stress-energy tensor for a fluid in an arbitrary spacetime with metric $g_{\mu\nu}$ with distinct radial and tangential pressures,
$p_r$ and $p_t$, is given by (e.g.~\cite{herrera}):
\begin{equation}
T_{\mu\nu} = (\rho + p_t)u_{\mu}u_{\nu} + p_tg_{\mu\nu} + p_t g_{\mu\nu} +(p_r - p_t)k_{\mu}k_{\nu}
\end{equation}
where $k^{\mu}$ is a unit four-vector orthogonal to the fluid's four-velocity $u^{\mu}$. The equations of motion are determined by 
$\nabla_{\mu}T^{\mu}_{\nu} = 0$; with the help of the projection tensor $\perp^{\nu}_{\mu} = \delta^{\nu}_{\rho} + u^{\nu}u_{\rho}$ 
($\delta^\mu_\nu$ is the usual Kronecker delta) these lead to the relativistic Euler equation:
\begin{equation}
(\rho + p_t)a_{\mu} = -\perp^{\nu}_{\mu}\nabla_{\nu}p_t - \perp^{\nu}_{\mu}\nabla_{\rho}(\sigma k^{\rho}k_{\nu})
\end{equation}
where $\sigma = p_r - p_t$ is the pressure anisotropy and $a_{\mu} = u^{\nu}\nabla_{\nu}u_{\mu}$ is the four-acceleration.

Considering the case of a static and spherically symmetric fluid star, the vectors defined earlier assume the following forms:
\bear
&& u^{\mu} = (u^t, 0, 0, 0)
\\
&& k^{\mu} = (0, k^r, 0, 0) 
\\
&& a^{\mu} = (a^t, a^r, 0, 0)
\eear 
The metric for this background system takes the familiar form $g_{\mu\nu} = \mbox{diag} (-e^{\nu(r)}, e^{\mu(r)}, r^2, r^2 \sin^2\theta)$.
The relativistic hydrostatic equilibrium is then governed by
\begin{equation}\label{StatSphEuler}
\partial_r p_r = -(\rho + p_r)\Gamma^t_{rt}-\frac{2\sigma}{r} 
\end{equation}
where $\Gamma^t_{rt} = (1/2)\, d\nu/dr$ is a Christoffel symbol. 

Perturbing the system away from this equilibrium induces a fluid flow $\delta u^\mu = (\delta u^t, \delta v^i)$ and a change 
in the density/pressure $\delta \rho, \delta p_r, \delta p_t$. In addition, we adopt the Cowling approximation where the perturbation in the
metric is ignored. At the same time, $k^\mu$ retains its unit magnitude and orthogonality to $u^\mu$, which
implies the following:
\begin{equation}
\delta k^r = \delta k^\theta = \delta k^\phi = 0, ~ ~ \delta k_t = -\frac{k_r}{u^t}\delta u^r
\end{equation}
With the help of these relations the perturbed Euler equation becomes,
\begin{eqnarray}
&a_{\mu}&(\delta\rho + \delta p_t) + (\rho + p_t)[\, u^{\mu}\nabla_{\nu}\delta u_{\mu} + \delta u^{\nu}\nabla_{\nu}u_{\mu} - u_{\mu}\delta u^{\nu}a_{\nu}\,]
\nonumber 
\\ &=& - \perp_{\mu}^{\nu}\nabla_{\nu}\,\delta p_t -\perp^{\nu}_{\mu}\nabla_{\rho}[\,\delta(\sigma k^{\rho}k_{\nu}\,]
\end{eqnarray} 

Decomposing this equation into its tangential and radial components, and simplifying each separately, we obtain:
\begin{equation}
(\rho + p_t)u^t\partial_t\delta v_i = -\nabla_i\delta p_t, ~~ i = \left\{\theta, \phi \right\}
\end{equation}
and
\begin{multline}
(\rho + p_t)u^t\partial_t\delta v_r - \sigma g_{rr} \frac{g^{tt}}{u^t}\partial_t\delta v^r = 
\\
= -\partial_r\delta p_r -\Gamma^t_{rt}(\delta \rho + \delta p_r) - \frac{2}{r}\delta \sigma
\end{multline}

To obtain the Newtonian limit of the linearised Euler equation, we assume $\rho \gg p_r, p_t$, $u_t \approx 1$, $g_{tt} \approx -(1+2\Phi)$ where $\Phi$ is the Newtonian potential. 
We then have ($i = \left\{r,\theta, \phi \right\}$)
\be
\rho\partial_t\delta v_i = -\partial_i\delta v_i - \delta^i_r\Big(\delta\rho\,\partial_r\Phi + \partial_r\delta\sigma + \frac{2}{r}\delta\sigma\Big) 
\ee
In this Newtonian expression we can easily add `by hand' the force due to the perturbed gravitational potential and the
viscous force $\nu\nabla^2\delta u_i$:
\begin{multline}
\label{SphLinEuler}
\rho\partial_t\delta v_i = -\partial_i\delta v_i -\rho\partial_i\delta \Phi 
\\
- \delta^i_r\Big(\delta\rho\,\partial_r \Phi  + \partial_r\delta\sigma + \frac{2}{r}\delta\sigma\Big)
+ \nu\nabla^2\delta v_i
\end{multline}
Writing this expression in Cartesian coordinates leads to eqn.~\eqref{fulleuler} which is the desired result.




\end{document}